\documentclass{article}
\usepackage{psfig}
\usepackage{emulateapj}
\usepackage{apjfonts}

\newcommand{\etal}{et~al.}

\lefthead{V\"ais\"anen et al.}
\righthead{Near-Infrared Galaxy Counts}

\begin{document}

\title{Wide-field $J$- and $K$-band galaxy counts in the {\em ELAIS} fields}

\submitted{Accepted to the Astrophysical Journal}

\author{P. V\"ais\"anen\altaffilmark{1,2},
E. V. Tollestrup\altaffilmark{1}, S. P. Willner\altaffilmark{1}, and
M. Cohen\altaffilmark{3,4}} 

\altaffiltext{1}{Harvard-Smithsonian Center for Astrophysics, 60 Garden St. 
MS 66, Cambridge, MA 02138}
\altaffiltext{2}{Observatory, P.O.B. 14, 00014 University of Helsinki,
Finland; present address; {\em vaisanen@astro.helsinki.fi}}
\altaffiltext{3}{Radio Astronomy Laboratory, 601 Campbell Hall,
University of California at Berkeley, Berkeley, CA 94720}
\altaffiltext{4}{Vanguard Research, Inc., Suite 204, 5321 Scotts Valley Drive,
Scotts Valley, CA 95066}

\begin{abstract}

New near-infrared galaxy counts in the $J$ and $K$ bands are
presented over a total area of 0.70 and 0.97 degrees$^{2}$,
respectively.  The limiting magnitudes of the deepest regions are
19.5 in $J$ and 18.0 in $K$.  At $J>16$ and $K>15$ our $J$ and
$K$-band counts number counts agree well with existing surveys {\em
provided} all data are corrected to a common magnitude scale.  There
are real differences from field to field, and the ELAIS N1 and N2
fields show an overdensity of $J<16$, $K<15$ galaxies.  The slopes of
$\log N(m) / dm$ are $\sim$ 0.40 to 0.45 at $15 < K < 18$ and $16 < J
< 19.5$.  Our counts favor galaxy models with a high normalization of
the local luminosity function and without strong evolution.

\end{abstract}


\section{Introduction}

Counting galaxies as a function of magnitude has established itself
as a powerful way to study galaxy evolution.  This is especially true
in the near-infrared, where K-corrections are smaller and dust
extinction of less importance than in the visible.  There have been
numerous $K$-band galaxy surveys for over a decade now. 
These have shown the pitfalls
of deriving any definite evolutionary conclusions from a single
wavelength: e.g.\ the excess of faint galaxies seen especially in blue
bands (see Ellis 1997 for a comprehensive review) is not nearly as
severe in $K$-band.

Another near-infrared waveband, the 1.25$\mu$m $J$-band, is now added
to the available data.  Bershady et al.\ (1998) and Saracco \etal\ (1999)
have published deep $J$-counts, and there are preliminary wide field DENIS- 
(Mamon 1998) and 2MASS-counts (Skrutskie 1999) available at
$J<16$.  Ultra-deep HST NICMOS counts are also available in $J$ and $H$ bands
(Thompson et al.\ 1999, Yan et al.\ 1998, Teplitz et al.\ 1998).
The only intermediate magnitude $J$-counts to date are those of 
Teplitz \etal\ (1999); our survey covers a slightly shallower magnitude range
but has a sky coverage 15 times as large.  

The range of magnitudes, $J$=12.5--19.5, in the present survey 
is too bright to be of direct interest to cosmology and high redshift
galaxy evolution studies.  However, modeling of the faintest counts
to study the evolution of distant galaxy populations heavily relies
on accurate normalization of the models at brighter magnitudes.
Indeed, the normalization is often left as a free parameter,
signifying that the local galaxy population is still quite
uncertain.  The bright and medium counts are thus instrumental
in the studies of both the local galaxy population and the evolving
properties of galaxies with large look-back times.

Our survey fields are within the European Large Area ISO-Survey
(ELAIS) regions (see Oliver et al.\ 2000).  ELAIS fields will be
heavily studied in wavelengths ranging from X-rays to radio bands,
with much of the scientific interest lying in high-redshift,
dust-obscured star formation, AGN, and ultraluminous infrared
sources.  Our observations were done as part of NIR followup for the
ELAIS project.  This paper presents our galaxy counts and compares
them to some available data and models.  We also address the question of
whether this field, selected because of its low 100~$\mu$m emission,
is a representative sample of the extragalactic
sky.  The discussion of individual objects relating to ISO-sources
will be presented elsewhere.

\section{Observations and data}

All the observations were carried out using the
STELIRCam instrument (Tollestrup \& Willner 2000) mounted on the
1.2-m telescope of Fred Lawrence Whipple Observatory on Mt.\ Hopkins.
The camera contains two InSb detector arrays with 256x256 format,
which view the same field simultaneously through different filters.  
Our $J$ and $K$-band images were taken during a total of 16 nights in
1997 April and June and 1998 April.  During the 1997 June run the
$J$-array was unavailable, resulting in larger coverage in $K$-band
in our survey.  The 1.2\arcsec\ pixel scale was used for all
observations, and the seeing varied between 1.8 and 2.3\arcsec. 

Each frame was a 60-second integration, typically consisting of 6 co-adds.  
A 5 by 5 dither pattern was used resulting in a 25 minute integration time per
pointing.  Data reduction followed a commonly-used sequence of linearizing,
sky-subtracting, flattening, and then co-adding the frames.  Cosmic rays 
were also removed in the process.  The partially overlapping 25 min 
pointings were then mosaiced together to produce the final images.
Near-infrared standard stars used to calibrate the data were taken
from Elias et al.\ (1982).  We estimate the calibration on the
instrumental system (Barr filters) to be accurate to 0.04~mag.

The survey consists of two areas located in the ELAIS N1 and N2
regions.  The central coordinates are ($\alpha, \delta$) = 16h 09m
00s, 54\arcdeg 40\arcmin 00\arcsec\ and 16h 36m 00s, 41\arcdeg
06\arcmin 00\arcsec\ (J2000), respectively.  The galactic latitudes
for N1 and N2 are $b$ = 45.0\arcdeg\ and 42.3\arcdeg.  During the
first run we mostly took data around ISO-detections. Later we filled 
in the gaps, but coverage of 
the N1 field remains fairly non-uniform.  The total 
areas observed are 0.70 degrees$^{2}$ in $J$ and 0.97 degrees$^{2}$ 
in $K$-band, less than half of the  ELAIS N1 and N2 areas, which
comprise $>2$~degrees$^{2}$.

We used the SExtractor v2.0.18 package (Bertin \& Arnouts 1996) for
the source detection and photometry.  At the deepest regions of our 
maps (see Table~\ref{tabcnts} for the areas)
the noise background level corresponds to $J=23.0$~mag/arcsec$^{2}$
and $K=21.3$~mag/arcsec$^{2}$ in $J$ and $K$, respectively.

SExtractor computed object magnitudes in several ways:
isophotal, various fixed apertures, and a composite referred to as
``BEST-magnitudes.'' The BEST-magnitude  
is usually the Kron-magnitude (see Bertin \& Arnouts 1996; Kron 1980), 
which is photometry in an  elliptical aperture with shape
determined by the  shape of the detected object and size chosen to
include almost all the object's flux.\footnote{The prescription for
computing Kron-magnitudes is, first, the second
order moments 
of the object profile are used to define a bivariate Gaussian profile
with mean standard deviation $\sigma_{\rm ISO}$.  
An elliptical aperture whose elongation $\epsilon$ and position angle 
are defined 
by these moments is scaled to 6$\sigma_{\rm ISO}$.  Next, within this
aperture the first moment is computed:
\[ r_{1} = \frac{\Sigma r I(r)} {\Sigma I(r)}. \]
Finally, the elliptical aperture used in actual 
photometry is defined by the
axes $\epsilon k r_{1}$ and  $k r_{1} / \epsilon$.  Additionally, the
smallest accessible aperture size is set to $3\sigma_{\rm ISO}$ to
avoid erroneously small apertures in the lowest $S/N$ regions. 
To arrive at a balance of systematic and random errors, a value 
$k\approx2$ is usually used.  We used $k=3$ which we found 
for our undersampled 
data to minimize the fraction of lost flux while still not increasing 
errors significantly.}
However, in very crowded regions the BEST-magnitude is
the isophotal magnitude instead.
In our data, this happens with fewer than 5\% of detections.
Based on our simulations using different detection parameters, 
corrections, tests on sub-fields, etc.,  the 
BEST-magnitude is the most robust.  It does not need aperture 
corrections, and thus there is no need to account for differences in  
flux measurements due to object profiles and sizes. 
Corrections for the differing seeing conditions during  
many observing nights are also rendered unimportant. 
(There is an input parameter in SExtractor for a Gaussian convolution 
used for {\em detection} of objects, but
it does {\em not} affect photometry.   Simulations showed that the
range of seeing observed had
$<5$\% effects on numbers of detections even at the faintest levels.)

Fig.~\ref{simul_avgs} shows the results of one of our simulations.
At each magnitude bin, $10^{4}$ objects were placed individually on
the real data frames and extracted in exactly the same way as the
final source catalog.  The simulations shown here are all for
galaxies, both disk and elliptical, because they best show the
complications in photometric measuring techniques.  
The aperture magnitudes lose a significant amount of flux when
measuring bright, extended galaxies.  Even large 17\arcsec\ apertures
typically underestimate the brightness of $K\approx13$ galaxies by
nearly 0.2 mag.  Small apertures, while being more robust at faint
levels, need very large (and often uncertain due to e.g.\ seeing
effects) corrections to total magnitudes at the bright levels.  The
pure isophotal magnitudes underestimate the flux at all magnitudes,
especially near the detection limit.  The faint {\em point} sources
are also a problem with large apertures.  While generally point sources
behave much better with all magnitude scales (typically flux losses
of $\sim 0.1$ mag),  the corrections to total  are very different
for them, complicating the task of placing the whole survey on a
consistent magnitude scale.
The BEST-magnitudes, in contrast, remain very robust over
the whole range of magnitudes and object types, and {\em no corrections are 
needed}.
  
Fig.~\ref{simul_avgs} also shows that all the magnitude 
systems used in the simulations significantly overestimate the
flux at the faintest level.  This upturn happens at $\sim 30 - 50$\% 
completeness limit and thus does not affect our final catalog.  (In the 
case shown, the catalog cutoff would have been the $K=17.25$ bin,
although the deepest 
parts go down to  bins at $J=19.25$, $K=17.75$.)
On the other hand,
{\em all} commonly used magnitude measuring systems suffer biases against
{\em low surface brightness} objects (Dalcanton 1997), and ours
is no exception.

We have compared the BEST-magnitude number counts against all the mentioned 
measuring techniques (along with appropriate corrections).  
All the differences remain at less than 10\% down to the catalog limit
($J=19.5$, $K=18.0$). 
The final uncertainties are larger than
this, so though the BEST magnitude is adopted for all subsequent discussion, 
the results would be about the same for aperture or isophotal magnitudes.

\section{Number counts}

\subsection{Completeness corrections}

The noise of our mosaics is non-uniform owing to using frames from 
three different runs, with different sky conditions, and to varying
exposure times from dithering and overlapping frames.  We thus
counted the objects in different bins according to the depth of
regions in the final images. (See Bershady et al.\ 1998 for a similar
approach.)  

Completeness corrections were
calculated using Monte Carlo simulations.  
As mentioned in the previous section, 
simulated sources were randomly placed onto
our real data frames and extracted in the same way as the final
catalog of sources.  This gave the completeness levels
as a function of magnitude for the depth reached in each region of the survey.
Fig.~\ref{simul_comp} 
shows an example of completeness levels in the deep regions of the
$J$-images using point sources and galaxy profiles.  
The simulation results gave a matrix of input/output
magnitudes.  This matrix contains not only  incompleteness at
a given magnitude level but also the amount of 'bin-jumping', i.e.,\ sources
being found at another flux level than their intrinsic magnitude.  The 
original counts can be calculated by inverting the matrix using the 
observed counts. (See e.g.\ Moustakas et al.\ 1997, Minezaki et al.\ 1998.)
We also have a powerful internal check for the completeness 
corrections: the corrected counts from the shallow regions of
our map should be consistent with the observed counts from the deepest
depth bins.  As 
an example, Fig.~\ref{shallow_deep} shows a case where the N2 $J$-band
counts from the deepest area of the map are plotted with counts
from a shallower region. The completeness-corrected
shallow counts do indeed match the deeper  data.

Completeness simulations were done with real data frames, so the corrections
include the effect of confusion, overlapping of sources.  
Depending on object morphology, the completeness limit would have been
overestimated by 0.25 -- 0.4 magnitudes if source-free noise frames
had been used in the simulations.

\subsection{Star/galaxy separation}

Separation of stars and galaxies was done in different ways for
different magnitude ranges.
For the brightest objects, stars and galaxies were easily separated by eye
with the SExtractor CLASS parameter used as a check.

At magnitudes $J > 15.5$, $K > 14.0$,
the spatial resolution of our infrared images is not good enough for
morphological separation of stars from galaxies to be secure.  Down
to 1.5 magnitudes fainter than 
these limits, we used a combination of visible classifications and
visible and infrared colors to classify each object as star or
galaxy.  Visible data came from the APS-catalog\footnote{The 
Automated Plate Scanner (APS)
databases are supported by the National Science Foundation, the
National Aeronautics and Space Administration, and the University of
Minnesota and are available at {\em http://aps.umn.edu/}.}
(Pennington \etal\ 1993), which lists magnitudes derived from Palomar
Observatory Sky Survey images.
Two sets of color-color plots were used for
both $J$ and $K$ sources to decrease uncertainties: 
$B-R$ vs.\ $B-K$ and $R-K$ and $B-R$ vs.\ $B-J$ and $R-J$.
Fig.~\ref{col-col} shows the resulting $B-R$ vs.\  $R-K$ plot for sources 
brighter than $K=15.5$, which is slightly {\em worse} than the other 
color-color separation for the $K$-band.  We show this to compare to 
equivalent plots by Saracco et al.\ (1997; their Fig.~3) and McCracken 
et al.\ (2000).  Saracco \etal\ found this classifier to fail
-- stars and galaxies completely overlap each other.  
It seems as if their galaxies are missing $K$-band flux.
The reason remains unclear, since
they use exactly the same aperture for optical and NIR bands.  Their aperture 
is quite small, though, only 5\arcsec.  Our plot, in contrast, shows 'total' 
magnitudes, which can be criticized by claiming that they introduce 
different samplings of source profiles in different bands.  However, we find
consistent results whether using either of the two sets of 
color-color plots, the APS-classification (which is morphological in nature),
or our eye-balling/SExtractor classification. In any case, we are not trying 
to produce exact color-color tracks for stars and galaxies here, only to 
separate them, and for that purpose our magnitudes and those of the 
APS-catalog seem to be very well suited.\footnote{When available, 
using $B-I$ vs.\ $I-K$ (eg.\ Huang et
al.\ 1997) will do an even better job in separating stars and galaxies.}

In the faintest bin where the color
method was used ($J=16.75$, $K=15.25$), the fraction
of $J$ or $K$ sources that do not have APS-correspondents is
$\approx$15\%.  The intrinsically reddest galaxies are the first to
be missing from the APS-list, but with our data it is impossible to quantify
exactly the proportions of non-matched stars and galaxies.  As the
correction is small, we simply divided the unidentified sources
between stars and galaxies, taking 2/3 to be galaxies in this bin and
half to be galaxies in brighter bins.  Even if all the missing sources
were classified as galaxies, the change in number counts would still
be well within the uncertainties of the final galaxy counts.

Beyond $J>17$, $K>15.5$, the number of missing sources grows rapidly.
Instead of attempting classification of individual objects, we
subtracted a model (Cohen 1994) of the Galactic point source
foreground from the total counts to give the galaxy counts. 
Fig.~\ref{starj} shows the resulting $J$-band star counts in the N2 field.  
The scatter is small and is 
similar  in the N1 field and in the $K$-band. 
The SKY model counts (Cohen 1994) used to derive galaxy counts 
for the faintest magnitudes
($J>17$, $K>15.5$) are also shown in Fig.~\ref{starj}.  The model
counts were
computed specifically for our two fields but were scaled by a factor
of 0.9 in order to better fit the star counts in the faintest bins
where direct separation worked. The same model has previously
been applied to galaxy counts by Minezaki et~al.\ (1998) and by Hall
\etal\ (1998).  The latter authors found a precedent for one
field (Q0736-063) with a chance under-density in foreground star
counts compared with SKY.  As is seen, the SKY model fits our star
counts very well.
 
\subsection{Count results and their uncertainties}

Table~\ref{tabcnts} and Figures~\ref{totalcntsj} 
and~\ref{totalcntsk} show our final galaxy counts.  The error
bars are calculated from Poisson statistics (Gehrels 1986).  In
addition, an uncertainty derived from the different star/galaxy
classifications and an assumed 10\% uncertainty of the SKY model are
included at bright and faint magnitudes, respectively.  In the four
faintest bins an estimated uncertainty of the completeness
corrections is also included.

The uncertainty from galaxy-galaxy correlations is negligible. Calculating  
equation 5 of Huang et~al.\ (1997) with our survey characteristics 
shows that at  most (at the faintest 
magnitude), the contribution from galaxy-galaxy correlations starts to 
approach 50\% of the Poisson uncertainties, which in turn are small
compared to  
our completeness correction uncertainties at these magnitudes.  
As noted by Huang et~al.\ (1997),
uncertainties due to large scale structures such as rich clusters and voids
are poorly known and hard to quantify.  Indeed, these might well be a major
reason for surprisingly varying results in galaxy counts
(Section~\ref{clusters}).

The effect of 
systematic magnitude errors ($\sim 0.05$ mag) on the final number
counts can be estimated from
equation 7 of Huang et~al.\ (1997).  The uncertainty
remains below Poisson uncertainties in our counts 
until $K=17$ and after that well 
below completeness-related uncertainties.  Photometric errors can
change the counts  
only in the horizontal direction, but random errors can affect also the log N 
slope. As seen in Fig.~\ref{simul_avgs}, photometric uncertainties
increase from  
0.02 at  $K=12$ to 0.5 mag at $K=17.5$.  Using Huang's equation 9 and a slope 
of 0.5 (Section~\ref{models}) this 
translates to an error of $\approx 0.01$ in the slope,
which is smaller than our derived 0.03 $1\sigma$ uncertainty in the slope 
fitting.  In any case, this effect resulting from galaxies jumping from bin 
to bin (and preferentially to a brighter magnitude bin near the detection 
limit) is corrected for in our completeness correction method.

Even if all corrections and the uneven depth of our
map were ignored, the derived counts would not change significantly
until the faintest two or three bins.  Treating the varying noise areas in
detail and modeling the bin-jumping gives us confidence to present
the corrected counts all the way to $J$=19.5 and $K$=18.




\section{Comparison with other data}

In $K$-band there have been numerous surveys for over a decade. 
Figure~\ref{totalk_comp} shows a compilation of available $K$ surveys 
at magnitude ranges similar to ours.
Our counts appear clearly and systematically higher than 
other surveys.  The difference is approximately a factor of two
at $K < 15$.  (At $K<13$ the statistical significance is small, but
at $13<K<15$ the difference is $>3\sigma$.)  
There is a factor of $\sim 1.4$ difference at $15 < K < 17$; the
most significant differences are with the Huang et al.\ (1997) and Gardner
et al.\ (1996) surveys due to their large areas ($\sim 10$ degrees$^{2}$).  
Non-uniform sky coverage is unlikely to explain the effect.  N2 was
covered very uniformly, while N1 was not, and $K$ more uniformly than
$J$, yet we see similar counts in both regions and bands.

There are two classes of explanation for the systematic differences.
The ELAIS fields may have galaxy populations that differ from those
in other survey fields, or different surveys may have used different
magnitude scales.  The latter would imply a 0.2-0.3 mag difference in
the fainter flux levels and 0.5 mag and more at $K < 15$.  The
following sections discuss each possibility.

\subsection{Galaxy Clusters and Local Overdensity}
\label{clusters}

Known galaxy clusters have a modest effect on our number counts.
The N2 region contains two galaxy clusters, Abell 2211 and 2213.  
Both are of richness 1, approximately 15\arcmin\ in diameter, and lie at 
a redshift of $z \approx 0.15$.  A 'typical' cluster member is 
expected to have $R \approx 18$, $J\approx 16$, and $K\approx 15$.   
We estimated the effect of these clusters on the counts by excluding the 
areas around them (16\arcmin\ diameter) 
and recalculating the galaxy counts in N2.
Compared to the original counts, we find correction factors ranging from 0.77
to 0.97 between $J$=14.25 and 
18.25, the largest correction being at $J=15.25$ as 
expected.  In $K$, the corrections are somewhat smaller (the $K$ 
sky coverage being larger) averaging a factor of $\sim 0.9$
between $K$=13.75 to 16.75.  The largest correction is 0.80 at $K=14.75$.  
The cluster-corrected counts are tabulated in Table~\ref{tabcnts}
and shown in Figs.~\ref{totalcntsj} and~\ref{totalcntsk}.

The N1 field has a cluster (Abell 2168) near the edge of the region.  
It lies at $z\approx0.06$ and thus covers a larger area on the sky. 
Whether some of its members (expected brightnesses $K<13$) 
are part of our N1 field is 
impossible to determine with the available data.  In any case the 
statistics are poor at $J<15$ and $K<14$, and we did not attempt corrections.
The same holds for a nearby cluster (Abell 2197; $z\approx0.03$) 
$1^\circ$ west of N2.  If there is contamination from this cluster, it would 
only be at $J\approx12$, $K\approx11$.  

The effect of field selection on prior counts cannot be ignored either.
Many surveys 
purposely avoid regions near clusters of galaxies.  While this is
understandable not to contaminate `field galaxy' counts, it may
bias the counts to voids and otherwise selectively 
underdense regions.  Most significantly, the largest surveys to date,
Gardner et al.\ (1996) and Huang et al.\ (1997; hereafter the Hawaii 
survey) both avoided clusters.

In order to try to quantify the effects of field selection,
we extracted blue magnitudes of galaxies from the fields of selected
$K$-surveys.  Fig.~\ref{apscnts} shows the blue 
(O~plate) APS-galaxy counts in the 
fields of the Hawaii survey, Gardner et al.\ (1996), 
Szokoly et al.\ (1998), and K\"ummel \& Wagner (2000),  
along with APS-counts in the regions of the present
survey.  All these surveys have $>0.6$ degrees$^{2}$ sky coverage.  
The APS counts are consistent with prior $B$-band counts 
in the fields of Gardner et al.\ (1996; using an 
approximate $m_{pg}-B=0.15$ color taken from Humphreys et al.\ 1991).

In the range $B$=18--20, which roughly corresponds to $K$=14--16, 
the N2 region shows an excess by a factor of 1.1--1.3 compared
to the other survey regions (including N1).  This excess in N2 is explained by
the clusters discussed above. The correction factors 
can be obtained from Table~\ref{tabcnts}. 

The fields of the Hawaii survey (Huang et al.\ 1997) show
systematically lower blue counts compared to other survey regions
by a factor of about 1.3. 
This survey contained numerous sub-fields (including the areas of
Glazebrook et al.\ 1994 counts), of
which we could examine 75\%. (The SB area was not available in the 
APS-catalog.)  There were large field-to-field variations, 
as pointed out by Huang et al.\ (1997). 
The lower amplitude of the 
Hawaii $K$-survey thus seems to be a result of a 
systematic {\em underdensity} at their survey regions, and
we adopt a factor of 1.3 correction to the 
Hawaii $K$-counts.  In fact, Huang et al.\  
interpret their counts (mainly the slope)
as pointing towards a large local void. 

The blue counts of the other large survey region (Gardner et al.\ 1996) are
at the same level as in our (corrected) fields.  
The authors have excluded a region around an unspecified cluster, and
thus the extracted  APS-counts from these fields 
are possibly slightly higher than the corresponding optical survey of 
Gardner et al.\ (1996).
There is a difference between their two fields, NGP and NEP; 
the NEP-field showed higher counts than the NGP field, even though the
cluster mentioned above is in the NGP field (Baugh et al.\ 1996).
Thus the known cluster cannot explain the difference between fields.
   
The other two surveys (K\"ummel \& Wagner 2000, Szokoly et al.\ 1998),
though smaller, show blue counts consistent  with those in our
regions and those of 
Gardner et al.\ (1996).   

All the above evidence shows that it is
practically impossible to define a pure `field galaxy' 
population when sky coverage is of the order of a square degree. 
There always are clusters, bright or faint, rich or poor, in or just outside
the field.  One can avoid the bright clusters, but never all the fainter
ones, so one necessarily expects a somewhat biased sample.
A search using NED\footnote{The NASA/IPAC Extragalactic Database (NED) is 
operated by the Jet Propulsion Laboratory, California Institute of Technology,
under contract with the National Aeronautics and Space Administration.} 
at the regions of the relatively uniform $K$-surveys in the literature
produced rich Abell
clusters within $1^{\circ}$ of nearly every field center.  
More distant ($z\sim0.5$)
rich clusters are expected to be numerous inside a 1 sq.degree field (see
eg.\ Lidman \& Peterson 1996).  For example, the SC field of Glazebrook et al.\
(1994) and Huang et al.\ (1997) has a rich Abell cluster just outside the
field, but more than 10 fainter galaxy clusters closer to the field center 
(NED, Lidman \& Peterson 1996).   The best one can do is to estimate 
the magnitude of the effect of either having clusters in field, or 
the effect of avoiding them.  Some surveys (Saracco 
et al.\ 1997, Ferreras et al.\ 1999) consist of tens of small, random
sub-fields.  In cases like these, it is easier to quantify the field-to-field
variations and, in fact, Saracco et al.\ (1997) find their systematically low
counts to be consistent with field-to-field count fluctuations.    
On the other hand, it is harder to accurately estimate the 
effective covered area due to large fraction of {\em edges} in the images.

In summary, there is a small ($\sim 10$\%) 
overdensity due to rich galaxy clusters in our N2 field.  
There is likewise
a systematic underdensity of galaxies in some of the Hawaii fields.
Additionally, there are nearby clusters outside both of our fields, 
which might affect the brightest ($J<14$, $K<13$) counts.

\subsection{Aperture corrections}
\label{apertures}

As mentioned above, the difference between our counts and
the others could also be that magnitude scales are not
directly comparable resulting in discrepancies in the horizontal
scale of the $\log$ N - mag plot.  We discuss this in the spirit that
all magnitudes {\em should be total}, since ultimately we wish to
compare data with models of galaxy populations which implicitly
assume total luminosities for the galaxies.  For the reverse
train of thought we refer the reader to an enlightening work by
Yoshii (1993), which models the photometric selection effects to
enable comparison of galaxy models with {\em raw} counts acquired 
with a given magnitude measuring method.

The Glazebrook et al.\ (1994) counts have been measured for the most part with
4\arcsec\ apertures.  Such an aperture is very small for galaxies in $13<K<16$ 
range and results in large corrections.  The authors present a correction to
physical, redshift dependent, $20 h^{-1}$ kpc apertures
(Glazebrook et al.\ 1995); the corrections range from $-1.0$ mag to $-0.1$ 
mag at 
$z=1.0$, typically being $-0.3$ to $-0.5$ 
mag around $K=15$.  Based on our simulations, we brightened their
counts by an additional $-0.4$ mag at $K$=14--16, and by $-0.5$ and
$-0.3$ at bins brighter 
and fainter than this range, respectively.  
Though the Glazebrook \etal\ counts
still remain somewhat lower than ours, the correction brings them
within $2\sigma$. Fig.~\ref{totalk_compcor} plots
these and other counts discussed below 
in their `corrected' form. 

\begin{figure}
\plotone{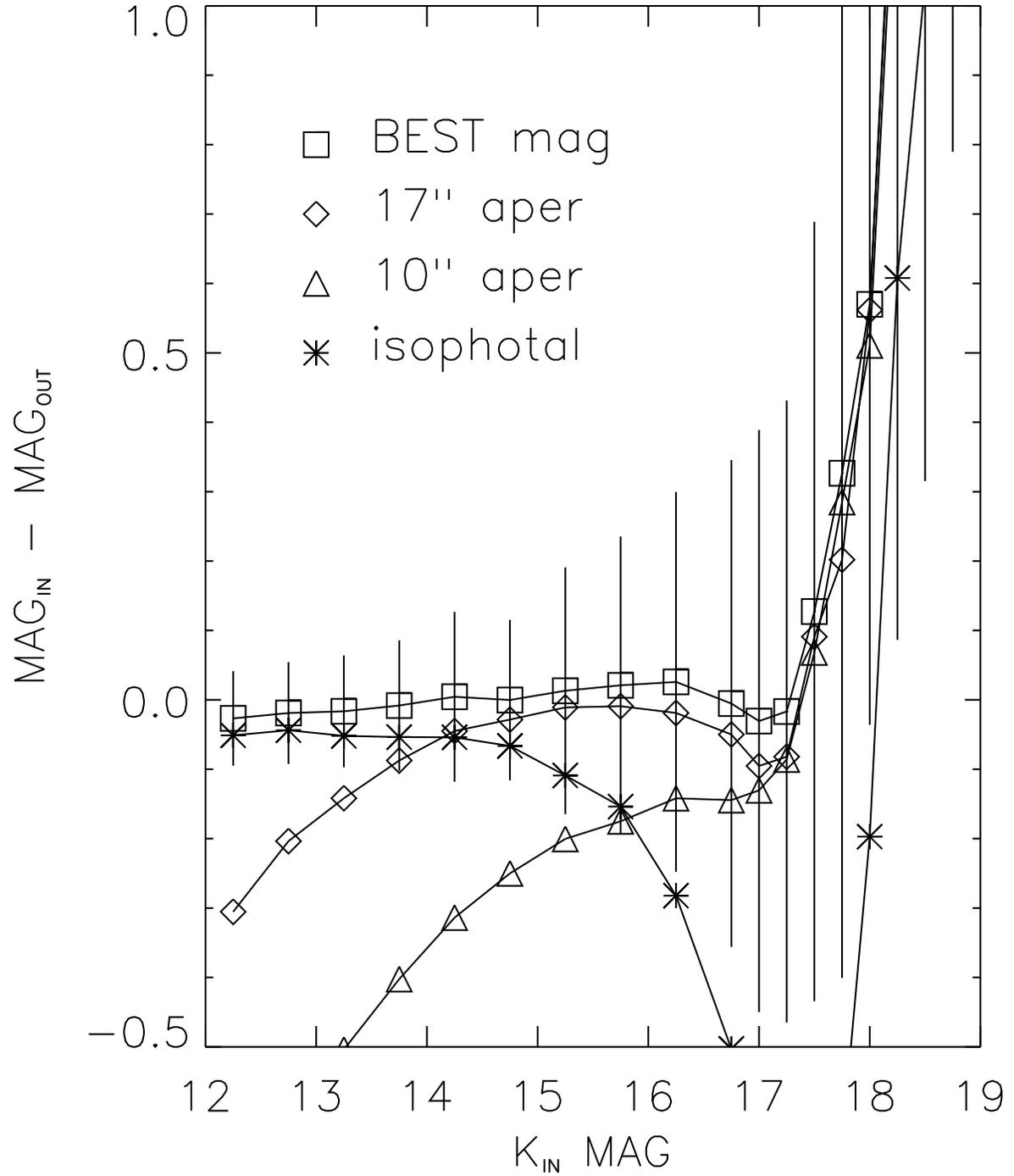}
\figcaption{Input minus output magnitude with different magnitude
measuring systems for simulated objects with galaxy profiles.  The
error bars are the standard deviation of fluxes of the detected
sources.  They are similar in all data sets and are plotted for only
one set for clarity.  These results are from Monte Carlo simulations
where $\sim 10^{4}$ sources in each bin were randomly placed in the
deepest 25\% of the K-band maps.  Images were
generated by IRAF's `MKOBJECTS' package and have random inclinations.
The BEST-magnitudes refer to
SExtractor output, which essentially uses the Kron-magnitude (see
text) to measure the flux.  In this case, the lowest bin in our
catalog would be the one at $K$=17.25~mag.
\label{simul_avgs}}
\end{figure}

\begin{figure}
\plotone{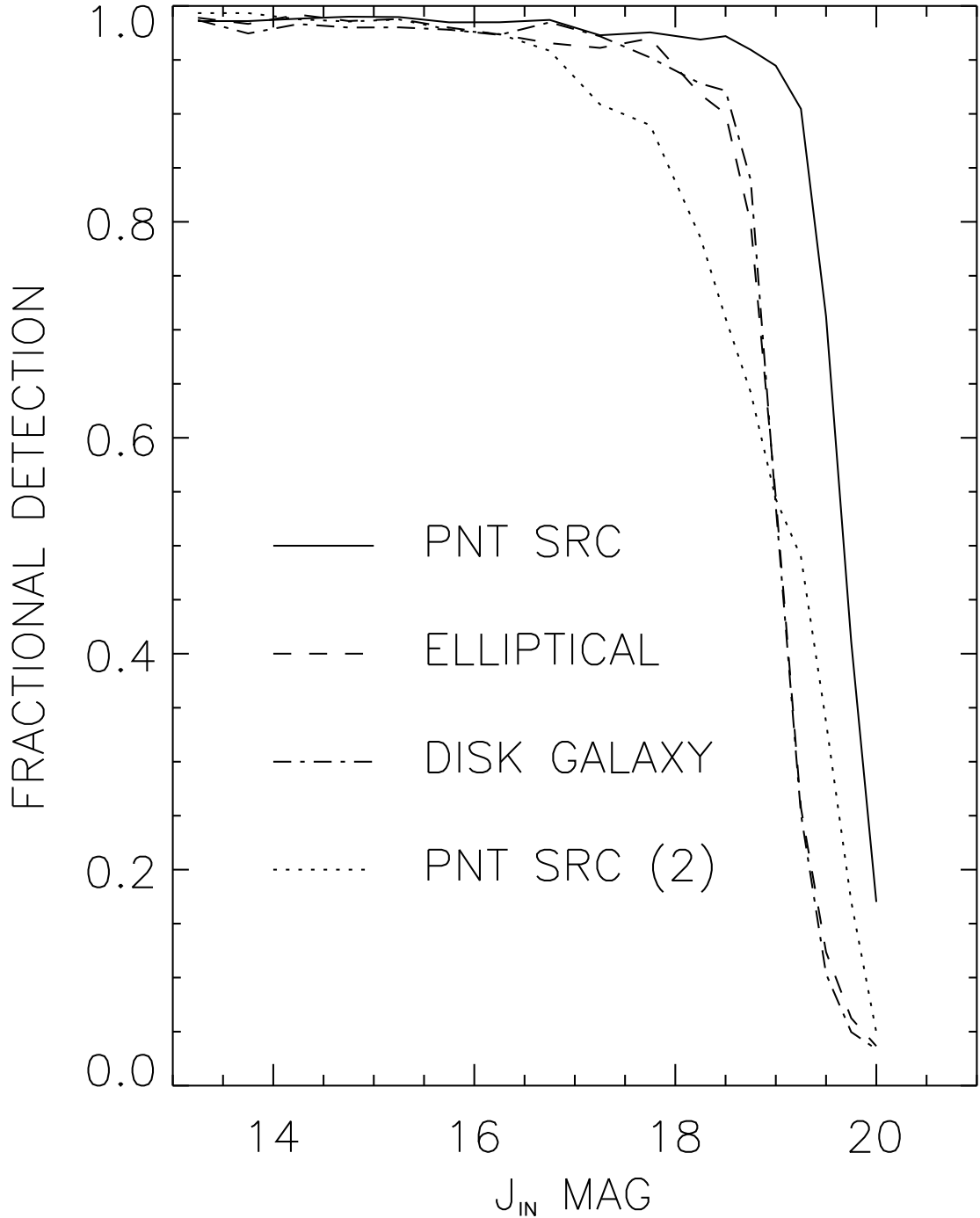}
\figcaption{Completeness levels from the photometric simulations.  Here
the fraction of detected sources in the deeper parts of $J$ maps are
shown for different object classes.  Point sources (solid line) are
detected to nearly 0.5 magnitudes fainter levels than normal galaxy
profiles (dashed and dot-dash lines).  The results shown
here use a simple definition for a `detection': the
extracted source was found within 1.5\arcsec\ of the input 
source and had a flux within $\pm4\sigma$ of the extracted spread of 
fluxes in the bin.  For the point source case a result using an
additional
requirement for detection (dotted line) is also shown: the extracted flux 
had to be within $\pm 0.25$ mag of input flux.  This definition would
allow a crude  completeness {\em correction}, but both of these
definitions overlook the issue of the derived magnitudes for sources
that are extracted but {\em not} within 0.25~mag of the correct flux.
For a better completeness correction, 
curves such as these are not enough.  One also 
needs the information on {\em where} the rest of the original sources
of the bin are extracted (`bin-jumping').  This information is
included in the Monte Carlo results.
\label{simul_comp}}
\end{figure}

\begin{figure}
\plotone{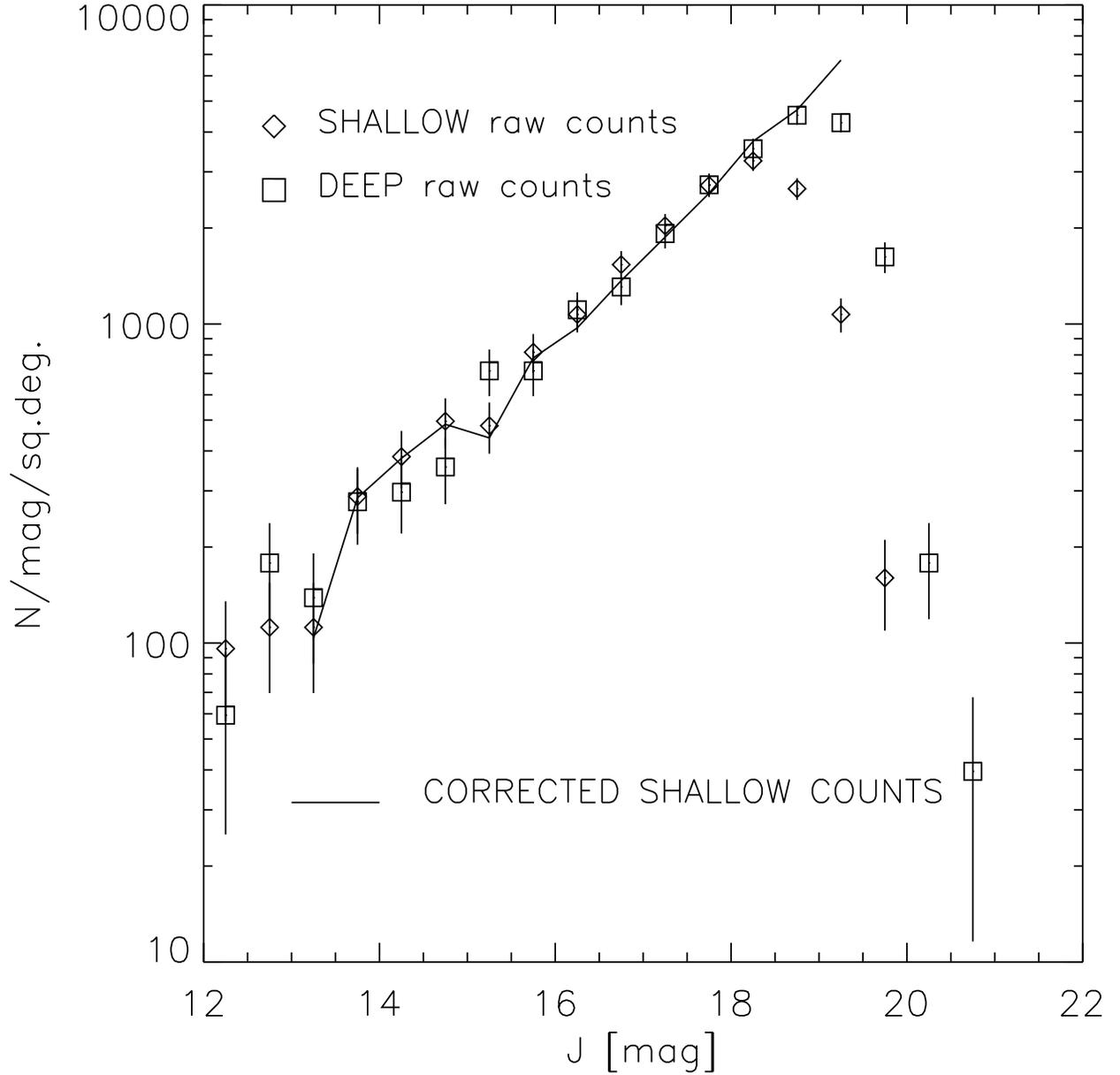}
\figcaption{$J$-band total counts in two different depth-bins of the
N2 area. 
The squares show raw counts from the deepest $\sim 30$\% of the map, and
the diamonds are from the second shallowest of the four different 
depth-bins ($\sim 25$\% of total area).  The solid curve shows the 
shallower counts after the completeness correction is applied.  Although 
the last point on the curve seems consistent, it has an effective
correction factor of about 7 and in practice is not
included in our catalog.  The corrected curve lies {\em below} 
the original raw 
count at some points because of bin-jumping effects: many faint sources
have erroneously been detected at the bin, and the correction moves them back
to their intrinsic, fainter bin.
The total (summed over the four depth bins) N2 $J$-band raw 
and corrected counts are shown in Fig.~\ref{starj}. 
\label{shallow_deep}}
\end{figure}

\begin{figure}
\plotone{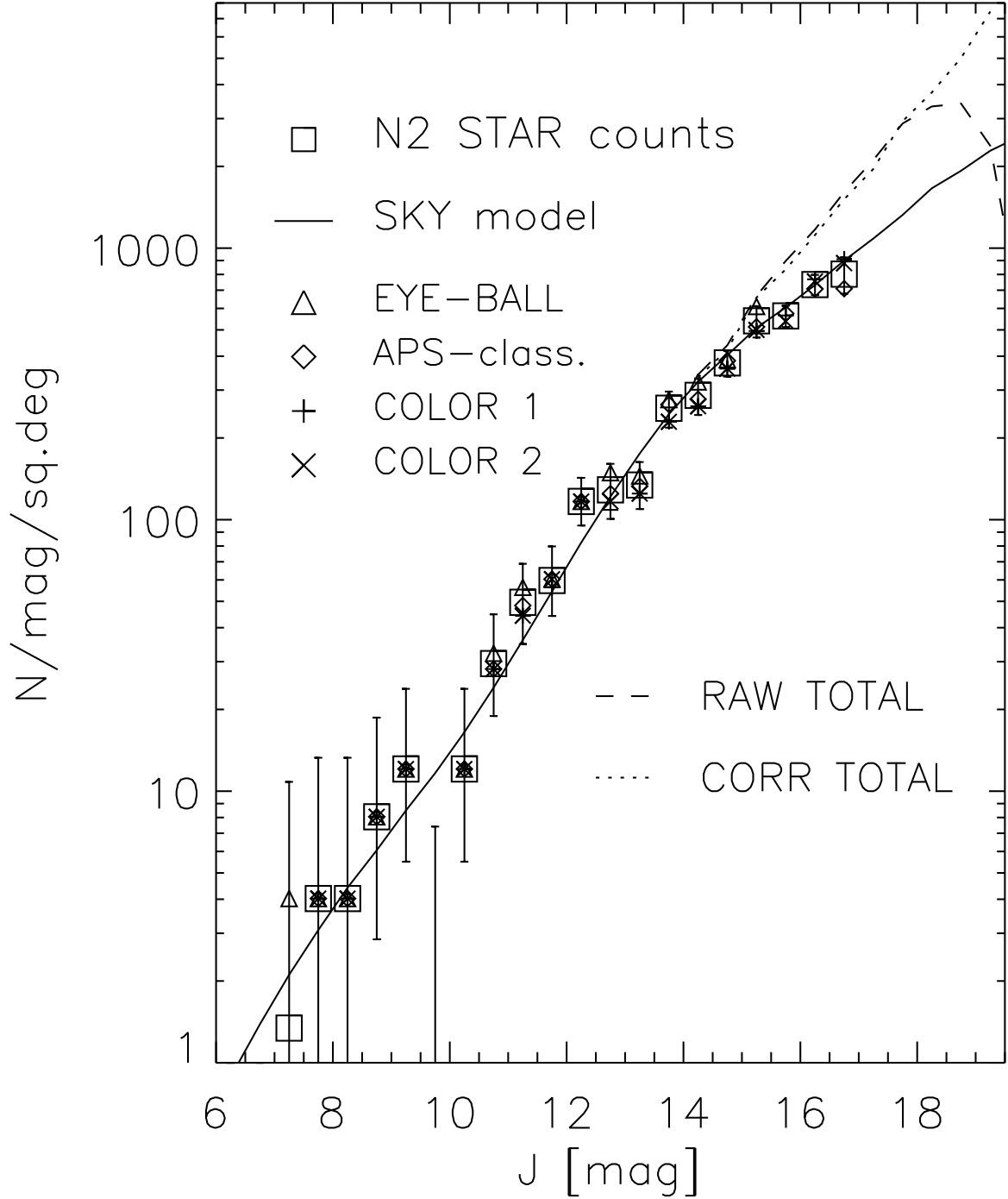}
\figcaption{Observed $J$-band star counts in the N2 field.
Squares show an average from different methods of classifying stars
versus galaxies:  eye-balling, using
the APS morphological classification, and
two different sets of color-color plots.
Color 1 refers to
$B-R$ vs.\ $B-J$ and color 2 to  $B-R$ vs.\ $R-J$.
The solid curve shows the SKY-model (Cohen 
1994) prediction for this field including the factor of 0.9 normalization.  
The {\em total} counts, stars plus galaxies,
are also shown as raw (dashed) and completeness-corrected 
(dotted).
\label{starj}}
\end{figure}

\begin{figure}
\plotone{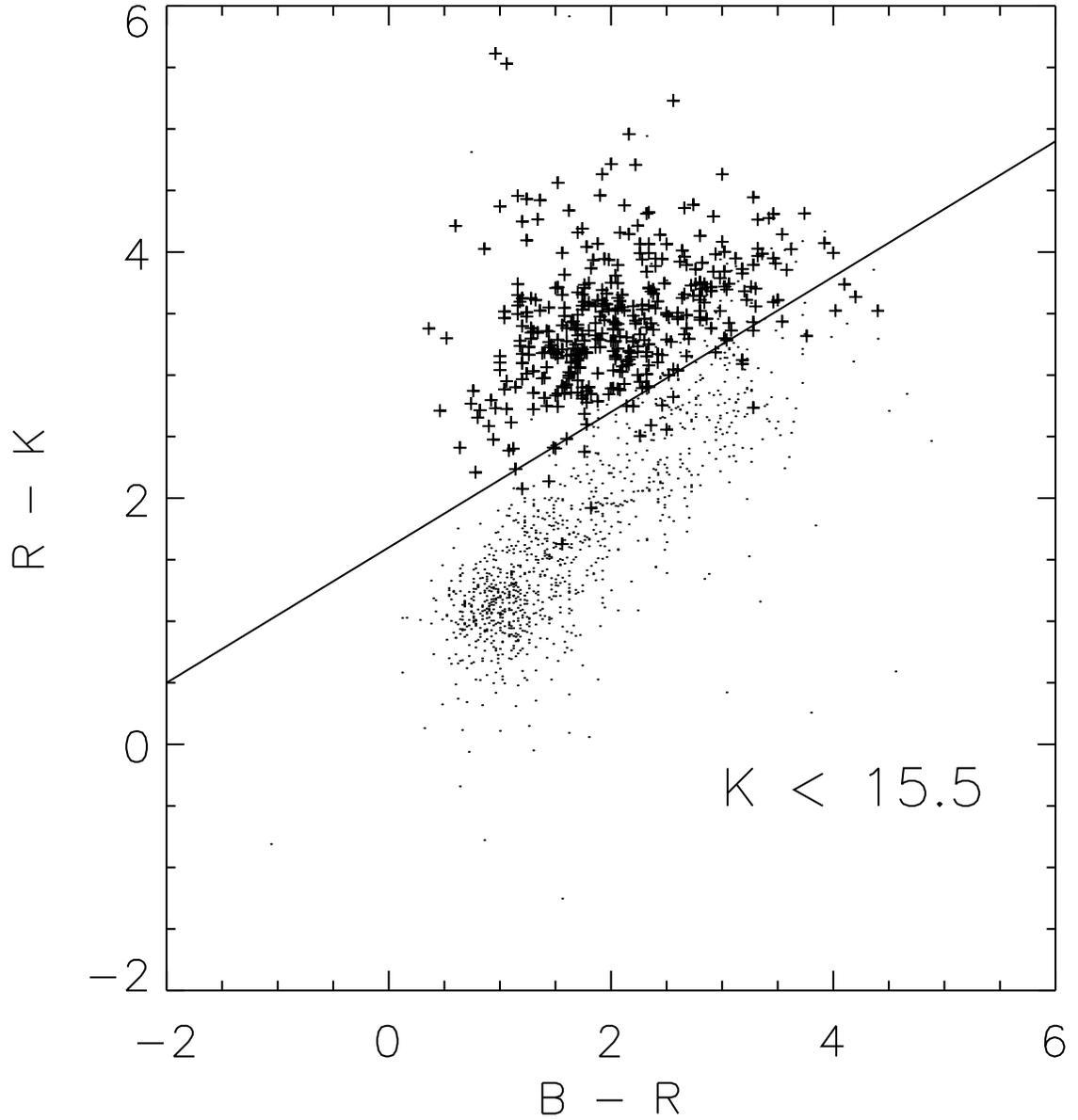}
\figcaption{An example of the color-color classification  for
the $K<15.5$ sources.  Objects separate into two groups in the $B-R$
/ $R-K$ plane, although at the red end there is more overlap.  The
classification is very consistent, however, with the morphological
classification of the APS-catalog: the crosses are galaxies and dots
stellar sources.  `$B$' and `$R$' in this plot actually mean the
POSS-plate blue (O) and red (E) magnitudes.  For the color-correction see
e.g.\ Humphreys et al. (1991).  \label{col-col}}
\end{figure}

\begin{figure}
\plotone{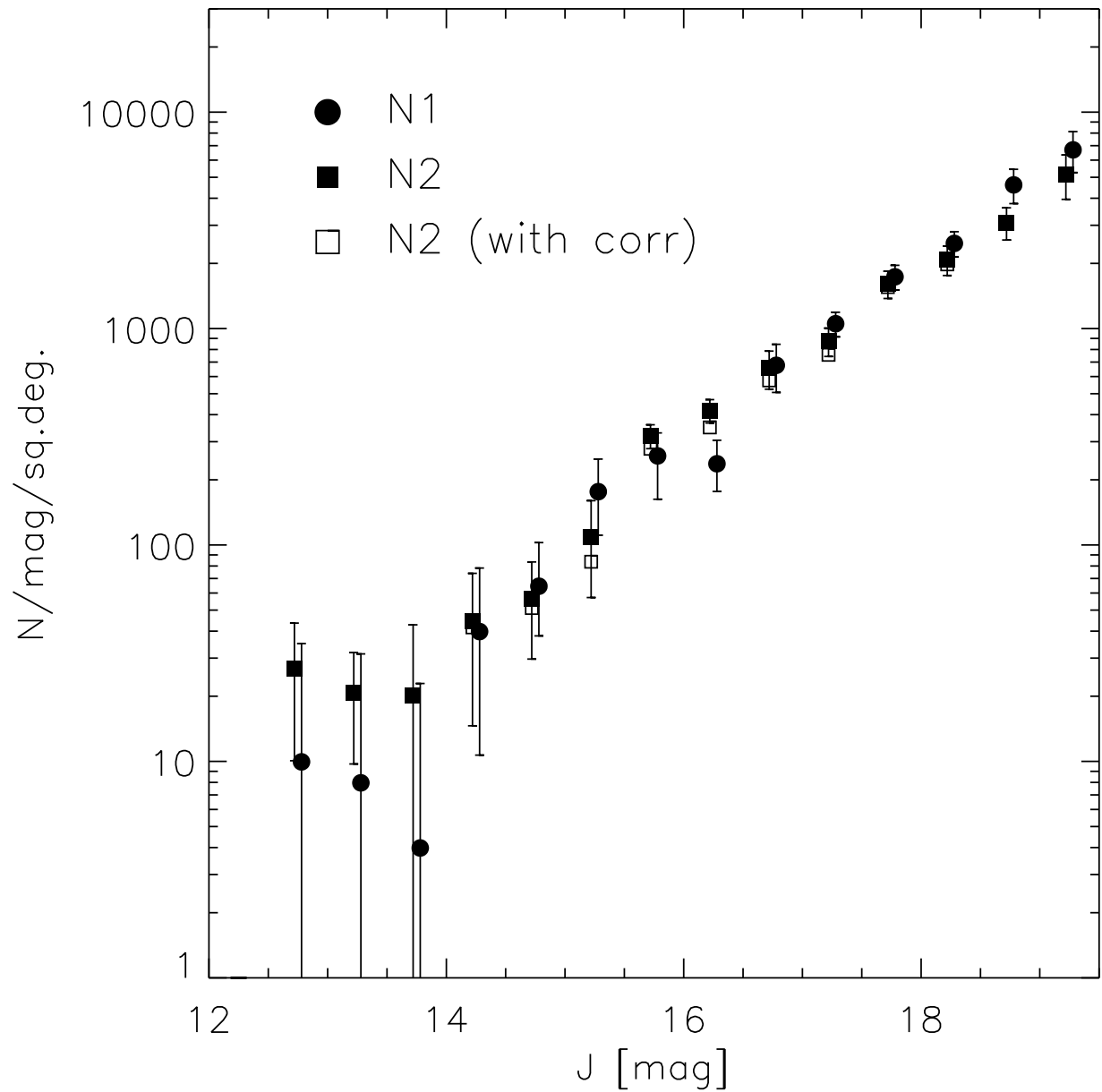}
\figcaption{Differential $J$-band galaxy counts.  
The counts are tabulated in Table~\ref{tabcnts}.  For counts in the
N2 region the open symbols show the effect of the cluster correction, 
discussed in Section~\ref{clusters}.  Error bars include the 
Poisson component and estimated completeness correction uncertainties.
The N1 and N2 points are offset by $0.03$ and $-0.03$ mag, respectively,
for clarity.
\label{totalcntsj}}
\end{figure}

\begin{figure}
\plotone{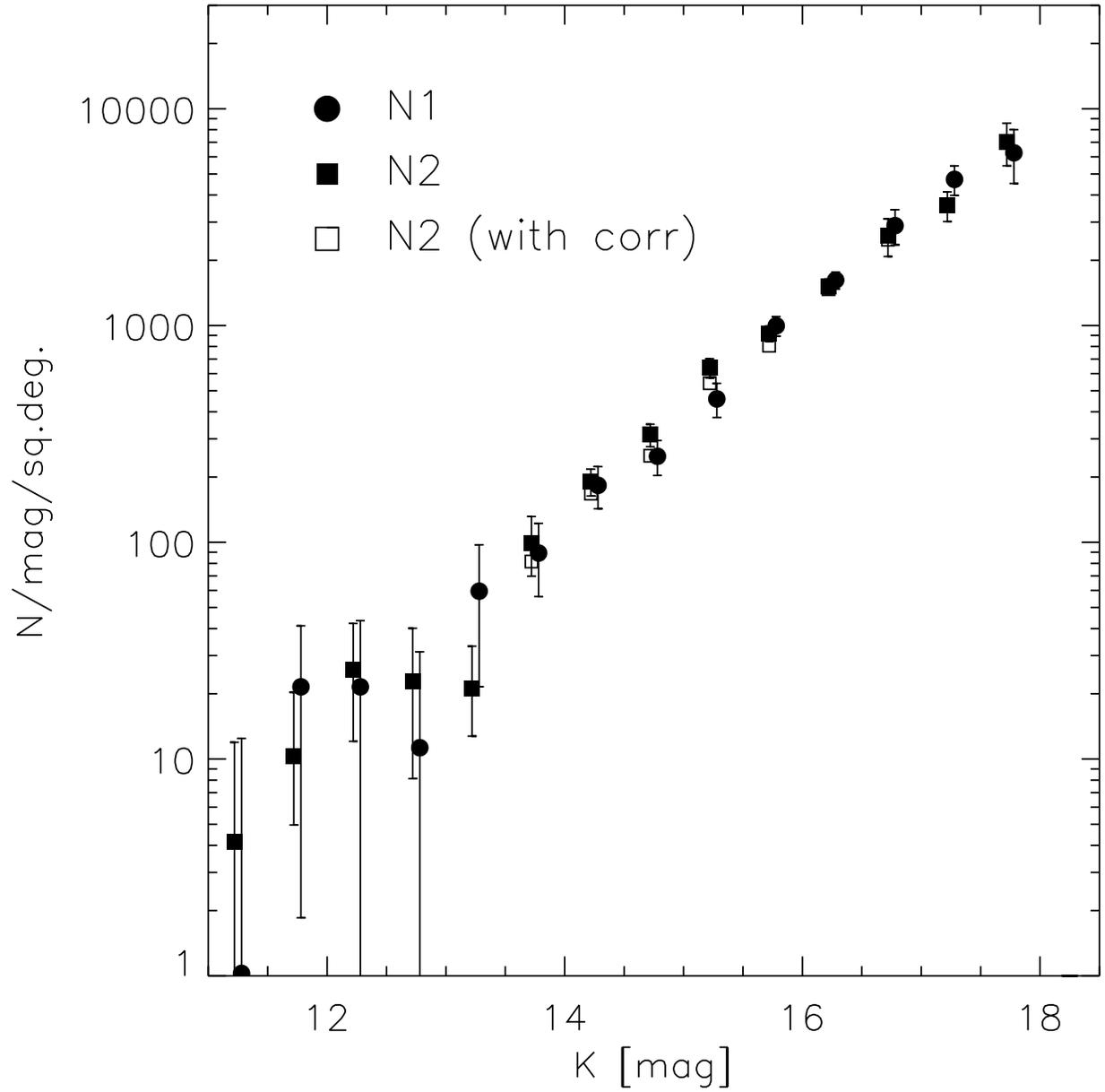}
\figcaption{Differential $K$-band galaxy counts.  
Open symbols show the cluster corrected counts, and the points are offset
by $\pm 0.03$ mag, as in the previous figure.
\label{totalcntsk}}
\end{figure}

\begin{figure}
\plotone{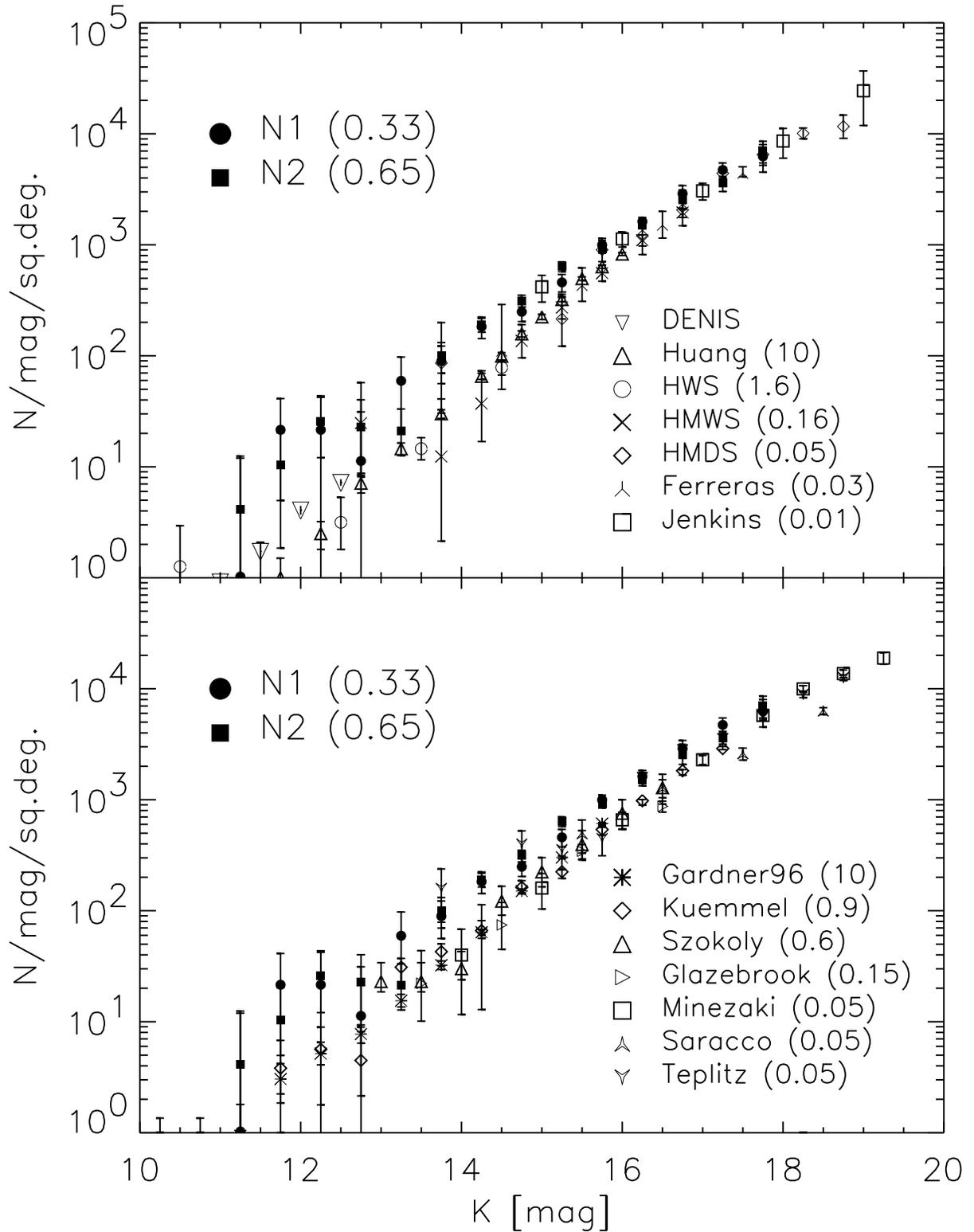}
\figcaption{$K$-band galaxy counts (solid symbols) along with other
available data at similar magnitude ranges.  All other data are plotted 
as published.  
Our counts show an excess compared to others, especially at bright 
magnitudes.
The comparison is split into two
panels for clarity only; 
there is no underlying difference between the two sets. 
To see the relative significance of the different counts, the approximate
sky coverages (in degrees$^{2}$) of the surveys are given in parentheses. 
The preliminary DENIS counts are a sample from all-southern-sky counts 
(Mamon 1998).  HWS, HMWS, and HMDS are from Gardner et al.\ (1993).
The rest of the surveys are: Jenkins \& Reid (1991), 
Glazebrook et al.\ (1994), Gardner et al.\ (1996),
Huang et al.\ (1997), Saracco et al.\ (1997), 
Minezaki et al.\ (1998), Szokoly et al.\ (1998),  
Ferreras et al.\ (1999), Teplitz et al.\ (1999), and 
K\"ummel \& Wagner (2000). 
\label{totalk_comp}}
\end{figure}

\begin{figure}
\plotone{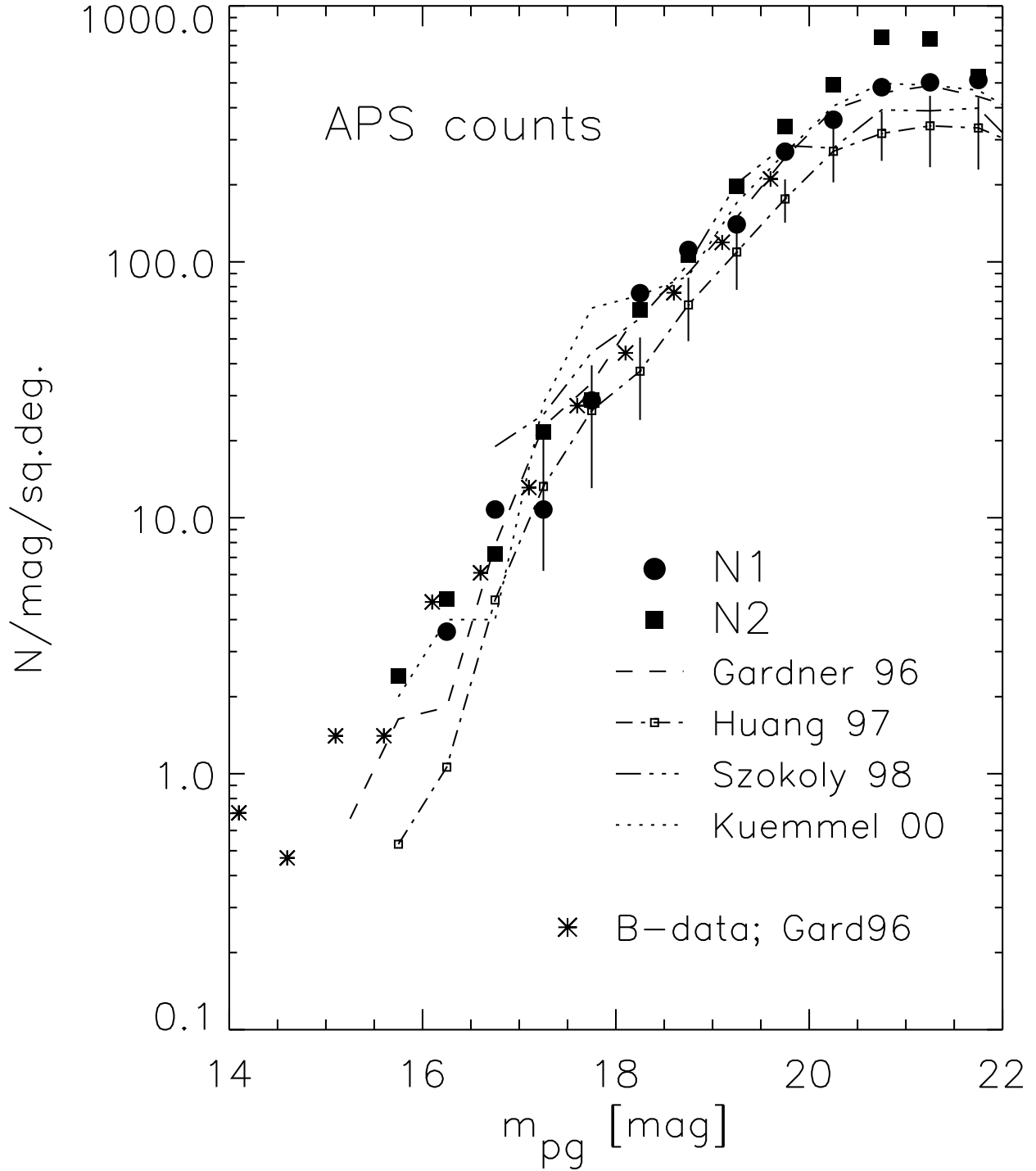}
\figcaption{Blue ($m_{pg} \approx B + 0.15$) galaxy counts extracted from the
APS-catalog (O plate).  Counts are shown at locations of selected large
$K$-band surveys, including our N1 and N2 fields.  
Counts at the areas of Huang et al.\ (1997) seem
to have a systematic underdensity of a factor $\sim1.3$ compared to
median counts.  Error bars for the Huang et al.\ (1997) counts
show the field-to-field 
variation.  The ELAIS N2 has an overdensity of about 10\%.  These
results are consistent with the $K$-band counts.  The
APS-based counts are not completeness corrected in any way, and the
incompleteness is expected to be significant at $B>20$.
\label{apscnts}}
\end{figure}

\begin{figure}
\plotone{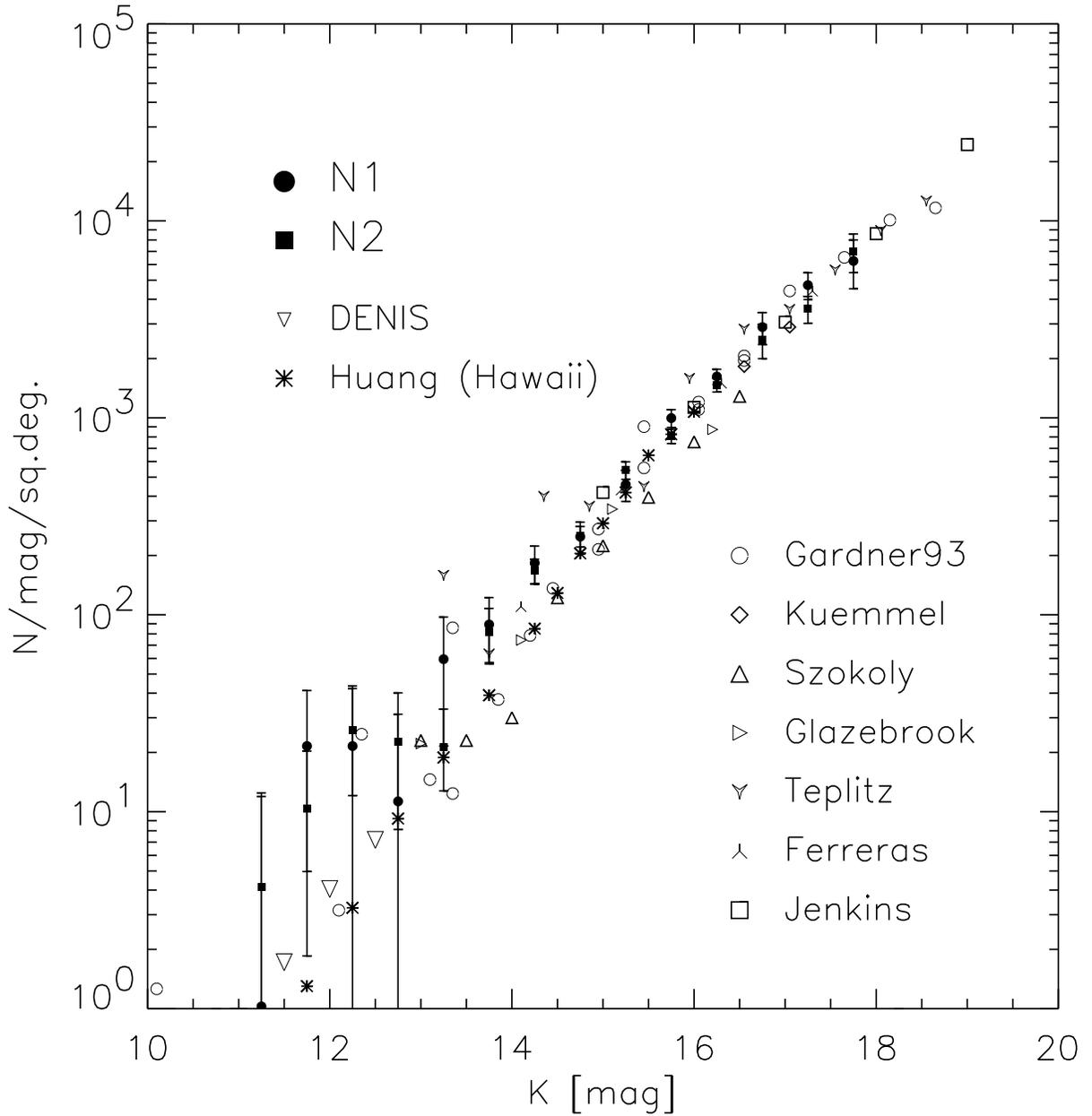}
\figcaption{Comparison of $K$ 
surveys that have magnitude systems directly comparable to ours 
plus those for which
we were able to estimate a magnitude correction. The latter are
plotted in their corrected form.
Data here are
a subset of those in Fig.~\ref{totalk_comp}. The 'Gardner93' label 
now contains the HWS, HMWS, and HMDS data.  Most error bars are 
omitted for clarity.
The Hawaii survey (Huang et al.\ 1997) is corrected 
for its systematic underdensity of galaxies, and our N2 counts are
corrected for their small 
overdensity (Section~\ref{clusters}, Figure~\ref{apscnts}), but
otherwise there are no field corrections. 
The $K>15$ counts are all consistent.  At
brighter magnitudes, while our statistics are poor, we seem to have an 
overdensity of galaxies. 
\label{totalk_compcor}}
\end{figure}

The Teplitz et al.\ (1999) counts are based on very small aperture 
magnitudes ($\sim 2$\arcsec), and we expect corrections of at least
$-0.5$ mag at $K$=14 decreasing to $-0.2$ mag at $K > 16.5$.  These 
corrections make their counts somewhat brighter (or higher)
than our counts, though still consistent.
The data of Gardner et al.\ (1993), Ferreras et al.\ 
(1999), and the $K>16.5$ bins of K\"ummel \& Wagner (2000) 
are also fixed aperture magnitudes.  The apertures range from 
6\arcsec to 8\arcsec. 
Based on our simulations (see eg.\ Fig.~\ref{simul_avgs}) 
even the 10\arcsec\ aperture underestimates the total flux of
galaxies by $\sim 0.4 - 0.2$ mag between $K=14-17$.  We corrected the 
Gardner et al.\ (1993) data by $-0.4$, $-0.3$, $-0.2$, and $-0.1$ mag at
bins $K<14.5$, $14.5 < K < 16$, $16 < K < 17.5$, and $K>17.5$.   The 
Ferreras et al.\ (1999) bins at $K=14.5$, $K$=15--17, and $K=17.5$ were 
corrected by $-0.4$, $-0.3$, and $-0.2$ mags, respectively.  Finally, the 
$K>16.5$ data of K\"ummel \& Wagner (2000) were corrected by $-0.2$ mag.

Gardner 
et al.\ (1996) used 10\arcsec\ apertures, but the results were corrected to 
'total' magnitudes using $I$-band growth curves.  However, the corrections 
and the equivalent aperture for the total magnitude are not available.

Saracco et al.\ (1997) and Minezaki et al.\ (1998), as well as 
K\"ummel \& Wagner (2000) in the brighter part of their magnitude range,
use FOCAS 'total magnitudes' (Jarvis \& Tyson 1981).  
All these authors note that the
FOCAS magnitudes tend to underestimate faint source fluxes (see also Thompson
et al.\ 1999).  The latter two groups thus apply additional
corrections: $-0.06$ to $-0.25$ mag at $17.5 < K < 19$  (Saracco et al.\ 1997)
and $\sim -0.1$ mag (Minezaki et al.\ 1998). The Saracco et al.\ (1997) 
counts lie below the bulk of other data, and the other two
are also fainter than our data by $\sim 0.3$ mag.  We have not explored 
FOCAS photometry and thus cannot quantify the exact differences involved here;
simple isophotal photometry, which the FOCAS total magnitudes are based on,
clearly underestimate the flux, as noted before (see  Fig.~\ref{simul_avgs} 
and also Saracco et al.\ 1999).

Huang et al.\ (1997) measure their magnitudes in 8\arcsec\ apertures, but 
correct them to 20\arcsec\ using curves of growth.  The corrections for 
galaxies are fairly large, ranging from $-0.55$ to $-0.2$ mag, 
consistent with the results of our simulations.  At $K<13$, Huang
\etal\ use isophotal
magnitudes for galaxies, which underestimate the total flux by 0.05 mag 
according to our simulations. Comparing the 20\arcsec\ fixed apertures to
our BEST-magnitude, we still find evidence of small 0.05--0.1
mag difference at $14<K<17$, part of which can also be due to a small 
overestimation of flux with BEST-magnitudes (Fig.~\ref{simul_avgs}). 
The cluster-corrected Hawaii counts are 
consistent with ours.  (See Section~\ref{clusters}; 
without the factor of 1.3 correction due to
underdensity, their counts would 
lie below ours with $\sim3\sigma$ significance). 

The counts of Szokoly et al.\ (1998) are the only ones which use exactly the 
same magnitude scale as we do.  Their counts also lie below ours with 
a difference of $\sim 0.3$ mag at $K>15$ (or a factor of about
1.3 in number), though they are not more than $1.5\sigma$ away from
our counts. 

Of the presently available counts the most consistent with ours 
without corrections
over the whole measured magnitude range 
are those of Jenkins \& Reid (1991).  Considering the 
observational technique, these counts are quite different in nature 
from the other counts.  They are a result of statistical evaluation
of the $K$-band background fluctuations in random patches of sky.  The method
has been widely used in radio and x-ray source counts (e.g.\ Condon 1974, 
Scheuer 1974) and should in 
principle be free of all the uncertainties arising from incompleteness in
detecting and measuring fluxes of individual sources.  Moreover, Jenkins \&
Reid (1991) do not purposely avoid clusters of galaxies, while many of
the other surveys do just that.  

Fig.~\ref{totalk_compcor} compiles those counts which had directly 
comparable magnitude systems or for which we have adequate information 
to make a magnitude correction.
Only those surveys using small apertures (Gardner et al.\ 1993, Glazebrook 
et al.\ 1994, Ferreras et al.\ 1999, Teplitz et al.\ 1999, 
and K\"ummel \& Wagner 2000 at $K>16.5$)
were corrected.  Jenkins \& Reid (1991), the Hawaii counts, and 
Szokoly et al.\ (1998) do not need photometric corrections compared to
our counts.
Most of the other surveys used FOCAS `total' magnitudes and we lack data to 
make a quantitative correction to the Kron-type magnitudes.  Though not 
plotted here for consistency, 
Saracco et al.\ (1997) and Minezaki et al.\ (1998) made corrections using
their own simulations.  Also Gardner et al.\ (1996) applied unspecified 
corrections to their aperture magnitudes.  After correction, all of
the counts are consistent for $K>15$.

There is much less data to compare with in the $J$-band 
(Fig.~\ref{totalcnts_modsj}).  The bright galaxy overdensity in both
N1 and N2 is
clearly seen also in $J$-band compared to DENIS counts.
At fainter $J$-magnitudes, our counts are in unison with the 
published Teplitz et al.\ (1999) counts and also connect very well with 
the deeper counts of Bershady \etal\ (1998).  
The most recent $J$-counts (Saracco \etal\ 1999) are consistent 
within the uncertainties, though they are clearly lower than e.g.\ the Teplitz 
et al.\ (1999) counts beyond our magnitude range. 
The counts of Teplitz et al.\ (1999) were measured with small 
($\sim 2\arcsec$) apertures, and after magnitude 
correction we would expect them to be somewhat brighter than our data, 
as the corresponding $K$-data.  

The Saracco \etal\ (1999) survey was measured with a 2.5\arcsec\ aperture.
Though the magnitude range is deeper than ours and small apertures are
thus justified, the size of aperture still might result in some flux-loss 
even with the fixed $-0.25$ mag aperture correction applied.  
Moreover, the field was centered on the 
NTT deep field, which possibly is 
biased against 'bright' ($J\sim 19$ mag here) objects.  

Bershady \etal\ (1998) also determined magnitudes using
small $\sim$ 2\arcsec\ apertures, but the measurements
were dynamically corrected to total magnitudes using the size of the object, 
in much the same manner as the Kron magnitudes we used.  Our counts
thus lie in the middle of the available $J>16$ data.

In summary, the differences among different sets of
$K$-counts at $K>15$ mag are
partially due to differences in magnitude scales.  After appropriate 
aperture and clustering corrections, 
our $J>16$, $K>15$ galaxy counts agree with those in the literature.  
Our bright ($K<15$, $J<16$) counts are still higher than most other data by
a factor of up to 2.  This is probably due to a real overdensity of 
bright field galaxies in our survey regions as compared to other
regions.  There are nearby clusters  
outside our fields, but it is hard to see how they could increase the 
counts by as much as a factor of two.

\section{Comparison with models}
\label{models}

Models incorporating cosmology and galaxy evolution should be able to
explain both the number counts themselves and the slope of the number
counts with magnitude.  Though a rigorous modeling of galaxy evolution
and galaxy populations requires knowledge of redshift distributions,
it is nonetheless informative to calculate how predictions from 
standard parameterizations of the local luminosity function (LF) fit
our final NIR counts. 

In a non-evolving model, the slope and amplitude of the counts are
dependant on the LF and the K-correction.  The latter depends on 
galaxy type mixes and SEDs of
galaxies, but these differences are small
in the NIR.  For a given LF, different cosmologies have
negligible effect on the counts in the bright range we consider.
Thus the only significant parameters are the local luminosity
function and luminosity evolution, which steepens the slope, $d \log 
N / dm$, at our magnitude range.  

To be sure we are not
affected by the bright galaxy excess in our fields, we 
examine ranges  $J$=16.5--19.5 and $K$=15.5--18.0.  All slopes, unless
otherwise stated, will refer to these magnitude ranges.  
Our $J$ counts show well determined slopes of 0.40$\pm$0.01 and
0.38$\pm$0.02  in N1 and N2, respectively 
(errors are $1\sigma$ uncertainties of the fitted gradient). 
In the $K$-band we  
measure slopes of 0.41$\pm$0.02 and 0.45$\pm$0.03.
The cluster correction in N2 steepened the slope slightly: without 
the correction, the $K$-slope would have been 0.43$\pm$0.02.   
Other recent counts, eg.\ Szokoly et al.\ (1998),
Minezaki et al.\ (1998), and K\"ummel \& Wagner (2000), give similar
values for the slopes in $K$.  
These are significantly shallower than the steep 0.65--0.70 
slopes at $K<16$ found in the Hawaii survey and by Gardner et al.\ (1996), 
even after taking into account the magnitude range difference. 
To explain the steep slope of the Hawaii
survey, Huang et al.\ (1997) invoked a significant under-density
of galaxies in the local universe at very large scales of over $300
h^{-1}$ Mpc.  Our data do not show evidence for this local hole
scenario. 

To model the counts, we first adopt the 
largest to-date NIR LF determination,
Gardner et al.\ (1997; hereafter Gardner LF), as the baseline model
($M_{K}^{\star} = -24.6$ mag, $\alpha=-0.9,
\varphi=2.1 \times 10^{-3}$ Mpc$^{-3}$, with $H_{0}=50$
km/s/Mpc).
We also make use of Gardner's (1998) galaxy number counts
software and use a basic parameter set found in the same paper, 
which includes six types of passively evolving galaxies with a galaxy 
mix from Gardner et al.\ (1997).
The Gardner LF with pure luminosity evolution results in a slope of
$\approx$0.46 in $J$-band and $\approx$0.48 in $K$-band.
No-evolution models give shallower slopes of 
$\approx$0.41 and $\approx$0.44, respectively. 
For reference, this same non-evolving 
model gives a slope of 0.60 at $K$=10--15, and 0.53 at $K$=13--18.  

The non-evolving Gardner LF model  best fits both our $J$ 
and $K$ slopes,
though the evolving slopes are within the $\sim3\sigma$ 
confidence limits.  
To see how model dependent the slopes are, we calculated
the relevant slopes for other observed LF's.  Mobasher et al.\ (1993),
Cowie et al.\ (1996),
Szokoly et al.\ (1998), and Loveday (2000) LF's
result in similar slopes as the Gardner LF, while the Glazebrook et al.\ 
(1995) LF gives slightly steeper slopes: 0.52 for the evolving model in $K$. 
Clearly, LF's resulting in steeper slopes than this would be 
ruled out by more than $3\sigma$ level by our counts.

\begin{figure}
\plotone{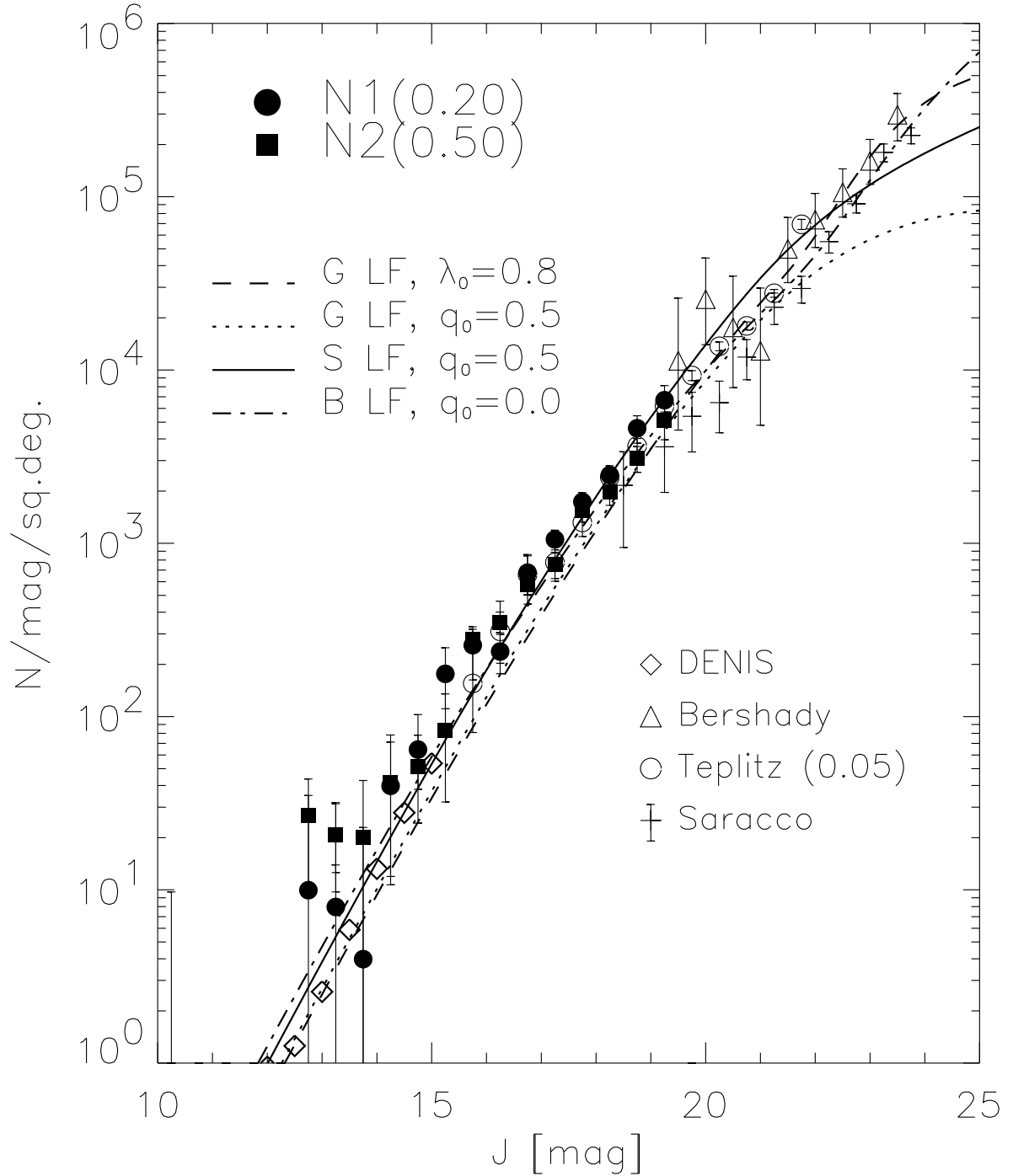}
\figcaption{Differential $J$-band galaxy counts.  
Our data are shown as solid symbols, circles for N1
and squares for N2 (cluster corrected).
Counts of Bershady et
al.\ (1998, triangles), Teplitz \etal\ (1999, open circles), and 
Saracco et al.\ (1999, crossed) are shown, along with 
preliminary counts from DENIS (Mamon 1998, diamonds).  Teplitz \etal\ (1999)
counts are the only ones done at a similar magnitude range, and their survey
area is indicated in parentheses.
All points are plotted as published.  We would expect
aperture corrections 
to make the Teplitz \etal\ (1999) counts
$\sim 0.2$ to 0.5~mag brighter.  Thus our counts would lie between
them and the Saracco et al.\ (1999) counts, while being most consistent
with the Bershady et al.\ (1998) data.   
There are several models plotted, all of which have
luminosity evolution included.  The lines labeled 'G' show
model counts using the LF of Gardner \etal\ (1997),
while 'S' is for Szokoly \etal\ (1998) LF, which has
more faint galaxies.  A model with a cosmological constant
($\lambda_{0}=0.8$) is shown as the dashed curve.  The
dash-dot line ('B') shows a model where the local LF 
parameters were chosen to fit an empirical simulation from 
Bershady et al.\ (1998; $1/V_{max}$, $q_{0}=0$) after which luminosity
evolution was added.  \label{totalcnts_modsj}}
\end{figure}

\begin{figure}
\plotone{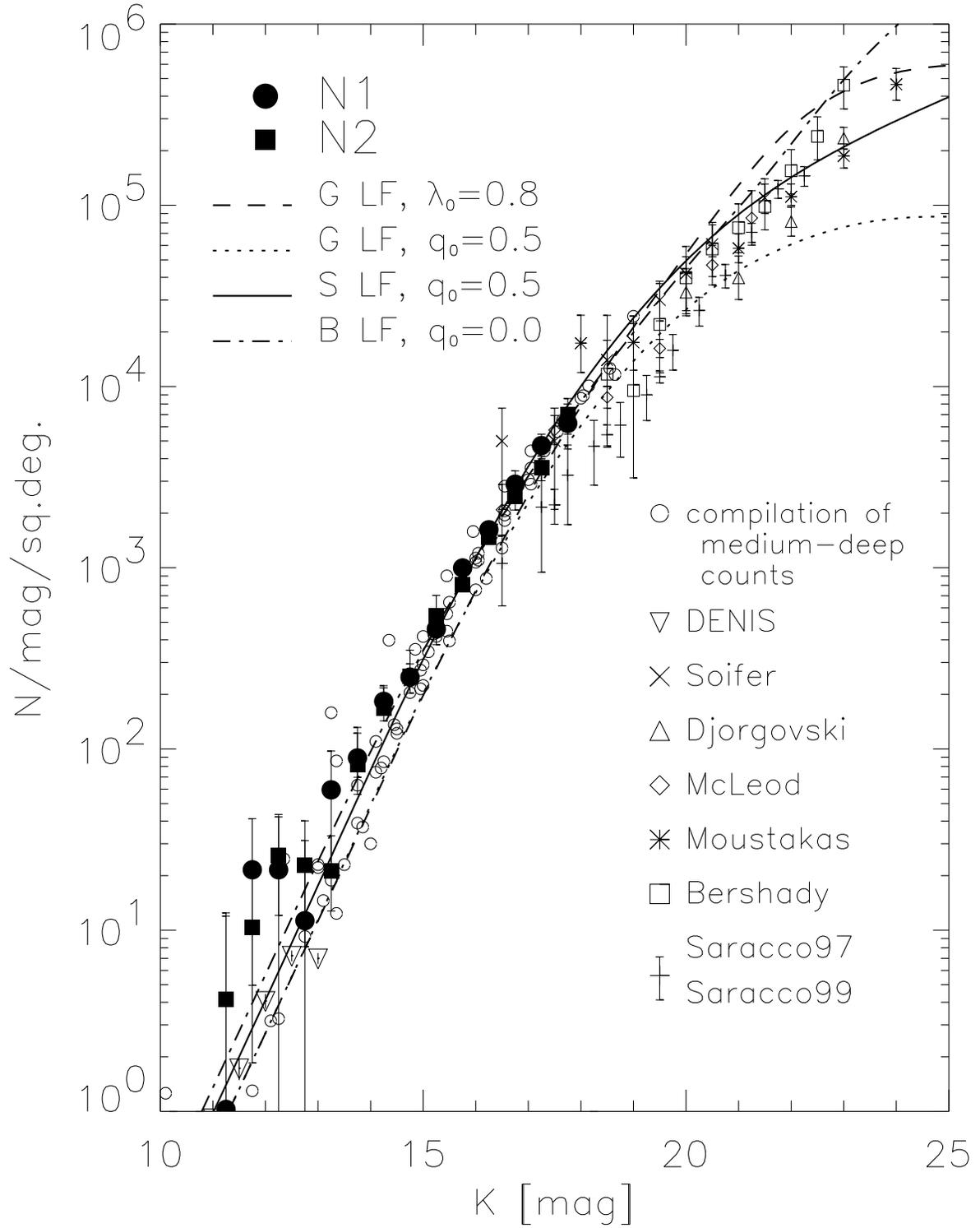}
\figcaption{Differential $K$-band galaxy counts.  
Our data are shown as solid symbols, circles for N1
and squares for N2 (cluster corrected).  Preliminary counts from DENIS 
(Mamon 1998) are shown as triangles.  Bright and medium deep 
counts are plotted here as open circles without separating them;
the same data-set are shown
individually in Fig.~\ref{totalk_compcor}.
The deep counts are shown separately. 
At $16 < K < 20$ the points below other 
data are those of Saracco et al.\ (1997, 1999); see discussion therein for
possible reasons for the underdensity.  Other deep survey data are from:
Soifer et al.\ (1994), Djorgovski et al.\ (1995), McLeod et al.\ (1995), 
Moustakas et al.\ (1997), Saracco et al.\ (1997; ESOKS1), and
Bershady et al.\ (1998).
Following Bershady et al.\ (1998), the Djorgovski et al.\ (1995) data are
corrected by -0.5 mag, all other deep data are plotted as published. 
The same models as in the previous $J$-band counts figure are
overplotted.  \label{totalcnts_modsk}}
\end{figure}

Examining both the slope and the amplitude of the counts
(see Figs.~\ref{totalcnts_modsj} 
and~\ref{totalcnts_modsk}), the baseline model underpredicts
our $J$ and the $K$ counts by more than $3\sigma$ at $13<K<17$.
This is easily understood, however, since the Gardner LF determination 
acquired the number density
$\varphi$ by fitting the counts of Gardner et al.\ (1996) and Huang et al.\
(1997).  These counts were seen (Sections~\ref{clusters} and~\ref{apertures})
to be lower than ours.  A factor of 1.5 higher value for the normalization
$\varphi$ in the Gardner LF
Schecter parameterization would give an excellent fit to our NIR counts.

In fact, the two most recent $K$-band LF determinations (Szokoly et
al.\ 1998, Loveday 2000) give direct observational support for
LF's resulting in a higher amplitude of number counts. (Both determined 
$\varphi$ by maximizing the likelihood of their Schecter parameters 
from their galaxy sample rather 
than fitting any given number count.)  While the 
exact value of number density $\varphi$ remains poorly constrained by all
present surveys, the best-fit values of both Szokoly et
al.\ (1998) and Loveday (2000) produce excellent fits to our counts
and have the same 
factor of 1.5 higher amplitude at $K<17$ compared to the baseline model.  
Fig.~\ref{totalcnts_modsj} and~\ref{totalcnts_modsk} show predicted counts
calculated from the Szokoly \etal\ (1998) LF 
($M_{K}^{\star} = -25.1$ mag, $\alpha=-1.3$, and $\varphi=1.5 \times
10^{-3}$ Mpc$^{-3}$).  This LF includes 
significantly more faint galaxies than the baseline model but also more 
$>L^{\star}$ galaxies.  The predicted counts from this LF, as well as that
of Loveday (2000), are  consistent with the fainter counts 
in both NIR bands when evolution is included.

By giving more freedom to LF 
parameters, it is possible to construct a realistic counts model which
produces the shallow slope ($\sim 0.4$)
of our NIR counts even with evolution included, along with
the correct amplitude, {\em and} a fit to the faintest counts.  
Bershady \etal\ (1998)
show counts from their observationally based ``$1/V_{max}$'' 
simulations.  We fitted a model to their $q_{0}=0$ curve 
($M_{K}^{\star} = -26.1$, $\alpha=-1.3$, and $\varphi=5.0 \times
10^{-4}$ Mpc$^{-3}$ gave a good fit) and added 
luminosity evolution.   
This LF has greater numbers of luminous galaxies than previously discussed 
LF's and also has more
faint galaxies than the baseline model (though not as many as the
Szokoly et al.\ 1998 LF).  While the fit to our own counts is convincing,
the LF seems to overpredict the preliminary DENIS counts.

As noted before, redshift distributions are essential in constraining
the models more accurately, e.g.\ to separate particular evolutionary 
models and different LF's producing similar number counts.  
There has been substantial progress in defining
IR-selected $N(z)$-samples (e.g.\ Cowie et al.\ 1996).  
The ELAIS regions are currently being 
followed-up with redshift-surveys, as well as multi-color imaging surveys;
there will thus be much improvement in the breadth of useful data 
in the near future.
We are in the process of getting our $J$-coverage on par with the $K$ survey,
and will also defer the discussion of $J-K$ color distributions to
a follow-up work.

In summary, non-evolving or passively 
evolving galaxy models best fit our NIR galaxy number count data.  
Models with stronger evolution or large local voids (which both 
result in steeper slopes) are ruled out by more than $3\sigma$ 
confidence.  The data favor local luminosity functions with 
relatively large populations of faint galaxies and those that produce 
a high normalization of mid-range NIR counts.

\section{Summary}

Our $J$ and $K$-band galaxy counts in 
two ELAIS fields (N1,N2) represent the largest areas
to date in the ranges $15 < J < 19.5$ and $16 < K < 18$.
The $J$-band counts are the first wide field galaxy counts 
at this magnitude range. 
For $J>16$, $K>15$, the data 
are consistent with existing surveys {\em provided} significant
magnitude scale corrections and large-scale 
structure effects are taken into account. 
In particular, the N2 region has a 10\%
overdensity of galaxies due to rich clusters.  The fields of the 
large Hawaii survey (Huang et al.\ 1997) 
seem to have a systematic underdensity of about a factor of 1.3, which
explains their counts without the need for a local void.   

At $J<16$ and $K<15$, the counts in the ELAIS fields are higher than
previous results by up to a factor of two.  This is probably due to a
real overdensity of field galaxies in the regions, although nearby
large galaxy clusters outside the fields may affect the very
brightest magnitude bins.  This overdensity needs to be taken into
account when interpreting results of surveys in other wavelengths in
these regions.

Our galaxy counts favor a high normalization of the local LF.  
The slope of the counts, $d \log N / dm \approx 0.40 - 0.45$, 
together with the amplitude of the counts at 
$K$=15--18 and $J$=16--19 are best fit by minimally
evolving galaxy models and luminosity functions having relatively large
numbers of faint galaxies.

\acknowledgments

We thank Gary Mamon for providing DENIS counts
and Matthew Bershady for providing his $J$-counts and models.  We also wish
to thank the referee and Kalevi Mattila for constructive criticism and 
useful suggestions. 
PV acknowledges support from the Smithsonian Institution, the Academy 
of Finland, and the Finnish Cultural Foundation.

\clearpage

\begin{deluxetable}{rrrrrrrrrrrrrrr}
\scriptsize  
\tablecolumns{15}  
\tablecaption{Differential galaxy counts\label{tabcnts}}
\tablewidth{0pt}
\tablehead{
\colhead{} & \multicolumn{6}{c}{N1} & \colhead{} & \multicolumn{7}{c}{N2} \nl
\cline{2-7}  \cline{9-15} \nl
\colhead{$mag$} & \colhead{$N_{raw}$\tablenotemark{a}} &
\colhead{$N$\tablenotemark{b}} & \colhead{$N_{low}$\tablenotemark{c}}
& 
\colhead{$N_{up}$\tablenotemark{d}} & \colhead{$A$\tablenotemark{e}}
& \colhead{$C$\tablenotemark{f}} & \colhead{} & \colhead{$N_{raw}$} &
\colhead{$N$} & 
\colhead{$N_{low}$} & 
\colhead{$N_{high}$} & \colhead{$N_{cc}$\tablenotemark{g}} &
\colhead{$A$} & \colhead{$C$} \nl  
}
\startdata
\multicolumn{14}{l}{$J$-band} \nl
12.75  &      1 &      10  &      0 &      35 &    0.201  &  1.000  & &      7  &     27   &    10   &    44 &  &   0.497 &   1.000 \nl
13.25  &      1 &       8  &      0 &      31 &    0.201  &  1.000  & &      5  &     21   &    10   &    32 &  &   0.497 &   1.000 \nl
13.75  &      0 &       4  &      0 &      23 &    0.201  &  1.000  & &      5  &     20   &     0   &    43 &  &   0.497 &   1.000 \nl
14.25  &      4 &      40  &     11 &      78 &    0.201  &  1.000  & &     11  &     44   &    15   &    74 & 42 & 0.497 &   1.000 \nl
14.75  &      6 &      64  &     38 &     103 &    0.201  &  1.000  & &     14  &     57   &    30   &    83 & 51 & 0.497 &   1.000 \nl
15.25  &     18 &     176  &    111 &     249 &    0.201  &  1.000  & &     27  &    109   &    57   &   160 & 84 & 0.497 &   1.000 \nl
15.75  &     26 &     258  &    163 &     329 &    0.201  &  1.000  & &     79  &    319   &   279   &   359 &278 & 0.497 &   1.000 \nl
16.25  &     24 &     237  &    177 &     305 &    0.201  &  1.000  & &    104  &    418   &   366   &   470 &349 & 0.497 &   1.000 \nl
16.75  &     68 &     676  &    507 &     846 &    0.201  &  1.000  & &    163  &    656   &   525   &   787 &577 & 0.497 &   1.000 \nl
17.25  &    113 &    1053  &    915 &    1190 &    0.201  &  1.068  & &    226  &    874   &   743   &  1004 &756 & 0.497 &   1.039 \nl
17.75  &    154 &    1734  &   1508 &    1960 &    0.201  &  0.885  & &    356  &   1609   &  1377   &  1841 &1555& 0.497 &   0.891 \nl
18.25  &    192 &    2475  &   2146 &    2804 &    0.201  &  0.772  & &    419  &   2083   &  1759   &  2406 &1979& 0.497 &   0.809 \nl
18.75  &    244 &    4611  &   3780 &    5443 &    0.142  &  0.747  & &    486  &   3091   &  2566   &  3615 &  &   0.417 &   0.754 \nl
19.25  &    136 &    6694  &   5260 &    8128 &    0.062  &  0.657  & &    390  &   5149   &  3955   &  6343 &  &   0.292 &   0.519 \nl
\cline{1-15} \nl 
\multicolumn{15}{l}{$K$-band} \nl
11.25  &      0 &       1  &      0 &      12 &    0.325  &  1.000  & &      1  &      4   &     0   &    12 &  &   0.645 &   1.000 \nl
11.75  &      4 &      22  &      2 &      41 &    0.325  &  1.000  & &      3  &     10   &     5   &    20 &  &   0.645 &   1.000 \nl
12.25  &      4 &      22  &      0 &      43 &    0.325  &  1.000  & &      8  &     26   &    12   &    42 &  &   0.645 &   1.000 \nl
12.75  &      2 &      11  &      0 &      31 &    0.325  &  1.000  & &      7  &     23   &     8   &    40 &  &   0.645 &   1.000 \nl
13.25  &     10 &      59  &     22 &      97 &    0.325  &  1.000  & &      7  &     21   &    13   &    33 &  &   0.645 &   1.000 \nl
13.75  &     15 &      89  &     56 &     122 &    0.325  &  1.000  & &     32  &     99   &    70   &   131 & 82 & 0.645 &   1.000 \nl
14.25  &     30 &     183  &    143 &     223 &    0.325  &  1.000  & &     61  &    190   &   163   &   217 & 168 & 0.645 &  1.000 \nl
14.75  &     41 &     249  &    203 &     295 &    0.325  &  1.000  & &    101  &    313   &   276   &   351 & 251 & 0.645 &  1.000 \nl
15.25  &     87 &     459  &    376 &     541 &    0.325  &  1.173  & &    204  &    638   &   572   &   703 & 542 & 0.645 &  0.991 \nl
15.75  &    168 &     997  &    894 &    1100 &    0.325  &  1.036  & &    311  &    917   &   842   &   993 & 807 & 0.645 &  1.051 \nl
16.25  &    238 &    1621  &   1476 &    1765 &    0.325  &  0.905  & &    475  &   1516   &  1394   &  1638 & 1417& 0.645 &  0.972 \nl
16.75  &    315 &    2889  &   2356 &    3422 &    0.325  &  0.670  & &    689  &   2593   &  2082   &  3103 &  &   0.645 &   0.824 \nl
17.25  &    301 &    4720  &   3987 &    5453 &    0.230  &  0.554  & &    719  &   3576   &  3017   &  4136 &  &   0.595 &   0.676 \nl
17.75  &    167 &    6260  &   4519 &    8001 &    0.122  &  0.436  & &    529  &   7019   &  5457   &  8580 &  &   0.458 &   0.329 \nl
\enddata
\tablenotetext{a}{actual number of galaxies counted}
\tablenotetext{b}{corrected counts in units of $N/mag/deg^{2}$}
\tablenotetext{c}{lower limit on corrected counts}
\tablenotetext{d}{upper limit on corrected counts}
\tablenotetext{e}{approximate survey area for this bin in square
degrees}
\tablenotetext{f}{approximate completeness for this bin}
\tablenotetext{g}{corrected galaxy counts after removing objects near clusters
for N2 region}
\tablenotetext{\null}{Note: Areas  and completeness
levels are approximations.  The exact values are determined
in four separate image depth bins, whereas only the global, averaged, 
values are shown here for comparative purposes.}
\end{deluxetable}

\clearpage

\end{document}